\documentclass[aps,prl,twocolumn,superscriptaddress]{revtex4-2}

\usepackage{amsmath,amssymb,graphicx}
\usepackage[english]{babel}
\usepackage[utf8]{inputenc}
\usepackage[T1]{fontenc}
\usepackage{amsmath}
\usepackage{mathtools,amssymb}
\usepackage{graphicx}
\usepackage{bm}
\usepackage{gensymb}
\usepackage{color}
\usepackage{hyperref}
\usepackage{upgreek}
\usepackage{scalerel}
\usepackage{pgffor}
\usepackage{pdfpages}
\usepackage{float}
\usepackage{appendix}
\usepackage{physics} 
\usepackage{lscape}   
\usepackage{silence}
\usepackage{soul,xcolor} 
\usepackage{comment}  
\makeatletter
\AtBeginDocument{\let\LS@rot\@undefined}
\makeatother

\begin{document}

\title{Bogoliubov flat bands in twisted layered materials}
\author{Keiji Yada}
\email[]{ yada.keiji.b8@f.mail.nagoya-u.ac.jp}
\affiliation{Department of Applied Physics, Nagoya University, 464-8603 Nagoya, Japan}

\author{Yuri Fukaya}
 \email[ ]{ yuri.fukaya@ec.okayama-u.ac.jp}
\affiliation{Faculty of Environmental Life, Natural Science and Technology, Okayama University, 700-8530 Okayama, Japan}

\author{Yukio Tanaka}
 \email[ ]{ ytanaka@nuap.nagoya-u.ac.jp}
\affiliation{Department of Applied Physics, Nagoya University, 464-8603 Nagoya, Japan}
\affiliation{Research Center for Crystalline Materials Engineering, Nagoya University, 464-8603 Nagoya, Japan}
\date{\today}
\begin{abstract}
Flat bands have attracted considerable interest in condensed matter physics because they provide a fertile platform for realizing strongly correlated and topological quantum phases. To date, however, most studies have focused on flat bands in normal-state electronic structures, such as those found in graphene and transition metal dichalcogenides. In this work, we investigate the emergence of flat bands in the superconducting Bogoliubov quasiparticle spectrum of twisted layered $d$-wave superconductors. We show that when the superconducting order parameter is odd under the in-plane $\mathrm{C}_2$ rotation, Bogoliubov flat bands can be engineered in the vicinity of the rotation axis. By analyzing a low-energy effective Hamiltonian, we demonstrate that the Berry connection of single layer system provides a clear criterion for the formation of the Bogoliubov flat bands. Our results establish a new paradigm of superconducting twistronics, in which the twist angle acts as a powerful tuning parameter for designing gapless flat-band superconductors.
\end{abstract}
\maketitle

The discovery of moir\'{e} superlattices in van der Waals heterostructures has opened a new frontier in condensed matter physics, enabling the exploration of flat electronic bands and correlated quantum phenomena \cite{Andrei2021,Balents2020}. 
Twisted bilayer graphene (TBG) first demonstrated that a small twist angle can dramatically reconstruct the electronic structure, producing nearly flat bands and giving rise to correlated insulators and unconventional superconductivity \cite{Lopes,
Bistritzer2011,Cao2018SC,Cao2018Correlated}. 
Similar moir\'{e} engineering principles have since been extended to transition metal dichalcogenides (TMDs), where strong interactions and tunable band structures enable realizations of Mott insulators, Wigner crystals, and generalized Hubbard models  
\cite{Tang2020,Regan2020}. These developments have established van der Waals moir\'{e} materials as a central platform in modern solid-state physics, contributing to the rapidly growing field of twistronics \cite{Andrei2021}.

A particularly intriguing aspect of moir\'{e} systems is the ability to engineer flat bands, which enhance interaction effects by quenching kinetic energy. While most studies have focused on flat bands in normal-state electronic structures, theoretical works have shown that flat or topologically nontrivial bands can also strongly influence superconducting properties, including superfluid weight and pairing mechanisms \cite{Kopnin2011,Volovik2019,Peotta2015,Heikkila2016,Julku2020,Isobe2018}. 
However, the possibility of realizing flat bands directly in the Bogoliubov quasiparticle spectrum of a superconductor remains largely unexplored.

The emergence of moir\'{e} superlattices in van der Waals materials has revealed that twisting layered quantum systems can fundamentally change their low-energy electronic structure. Interestingly, a conceptual parallel exists between graphene and $d$-wave superconductors: both host Dirac-like quasiparticles at low energies\cite{Ryu2002}. In graphene, the linear dispersion arises near its Dirac points, whereas in $d$-wave superconductors, nodal quasiparticles exhibit an analogous dispersion near the gap nodes \cite{LeeNagaosa,Randeria}. These Dirac points and nodes are protected by topological winding numbers \cite{SatoPRB2011}, highlighting a shared topological character between the two systems. This analogy naturally raises the question of whether phenomena discovered in moir\'{e} graphene—such as flat bands, strong correlations, and emergent topological phases—might also arise in twisted $d$-wave superconductors. 

Recent theoretical proposals and experimental advances have begun to explore this direction, marking the early development of what has been termed ''cuprate twistronics'' \cite{Confalone2024}. Phase-sensitive experiments on cuprate superconductors revealed remarkable signatures stemming from 
the sign-change of $d$-wave order parameter \cite{Tanaka95,Kashiwaya_2000,HuPRL1994,TsueiRMP,VanHarlingenPRL,SatoPRB2011}, providing a natural foundation for investigating twist-induced interference and frustration effects. Within this context, twisted bilayers of nodal superconductors offer a unique platform to study how interlayer phase structure and twist angle reshape the Bogoliubov quasiparticle spectrum. Recent theoretical works have predicted a variety of twist-induced phenomena, including magic-angle flat Bogoliubov bands \cite{Volkov2023}, chiral or topological superconducting phases emerging from large-angle twists \cite{Liu2023}, and even high-temperature Majorana modes in twisted cuprate heterostructures \cite{LiLiu2023,Mercado2022}. Experimental proposals such as twisted Josephson interferometry further suggest feasible routes to probe the symmetry and topology of these engineered superconducting states \cite{Berg}.
Thus, it is very timely to elucidate the energy dispersion and origin of the possible Bogoliubov flat band. 

In this work, we study the energy spectrum in twisting bilayer systems composed of nodal $d$-wave superconductors. By solving the Bogoliubov de Gennes (BdG) equation, 
we find the flattening of the quasiparticle band which we refer to as Bogoliubov flat band.
From the analysis of $4 \times 4$ effective Hamiltonian, the condition to realize the Bogoliubov flat band is clarified. 
%

\begin{figure}[t!]
    \centering
    \includegraphics[width=8.6cm]{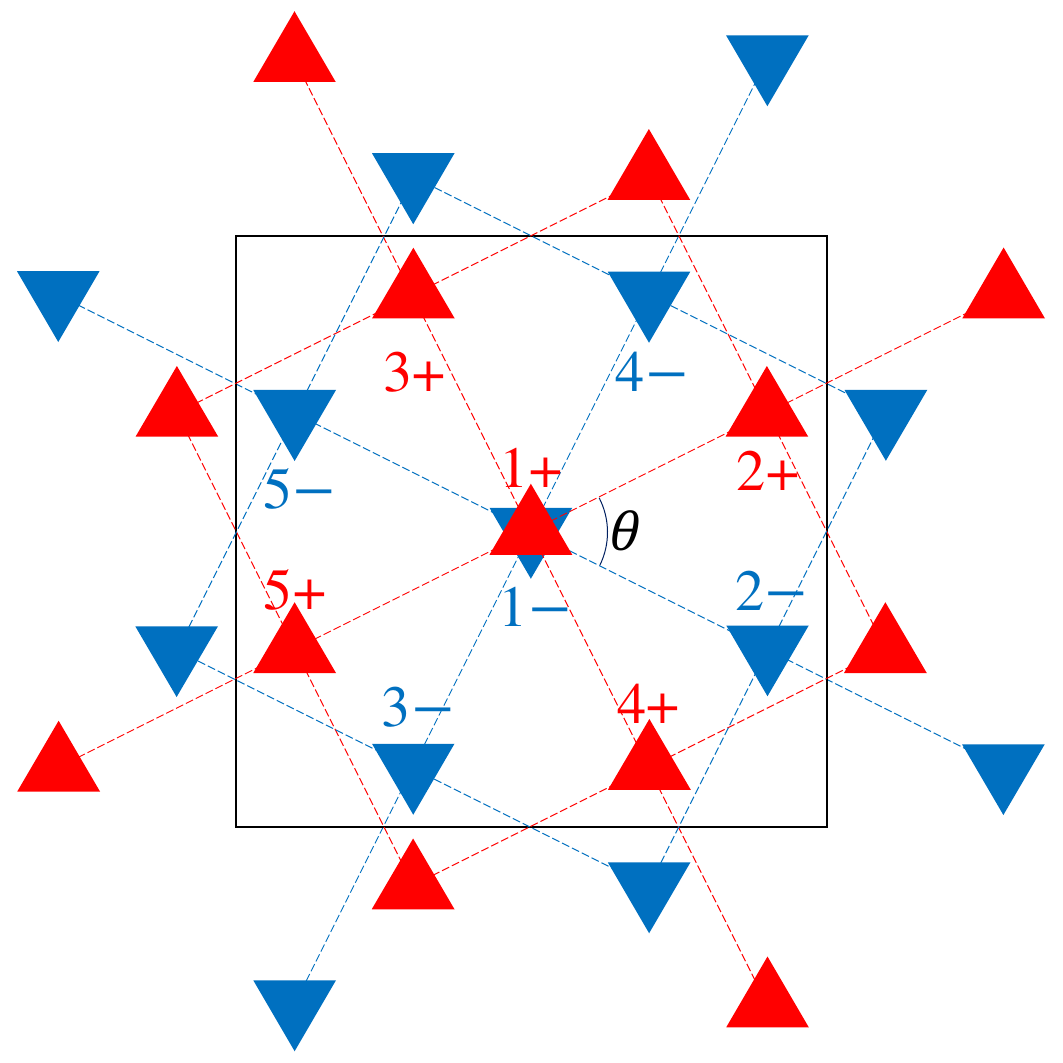}
    \caption{
    Structure of twisted bilayer square lattice.
    Regular and inverted triangles denote lattice points on the top and bottom layers, respectively. Squares with a solid and dotted outlines represent the unit cell of bilayer and single-layer systems, respectively. There are five atoms per unit cell in each layer.}
    \label{fig:lattice}
\end{figure}

\textit{Model and Formulation}.---In this section, we introduce a model for twisted layered structure and formulation to see the quasiparticle excitation.
We consider a bilayer structure composed of two-dimensional square lattices. Without twisting, the lattice points are given by $(ma, na,\pm c/2)$ where $a$ is in-plane lattice constant, $c$ is distance between the layers,
$m$ and $n$ are integers.
Next, we consider the twisting by the rotation of each layer along the $z$-axis.
The rotation angles of the top and bottom layers are given by $\theta_+$ and $\theta_-$, respectively. Then, the lattice points for the twisted layered structure are given by $(ma\cos\theta_\pm +na\sin\theta_\pm , na\cos\theta_\pm-\sin\theta_\pm,\pm c/2)$. Since only the relative angle between the layers $\theta$ is relevant, we set $\theta_\pm=\pm \theta/2$. When the value of $\tan(\theta/2)$ is a rational number, the layered structure has periodicity. The simplest case is $\tan(\theta/2)=1/2$ whose lattice constant is $b=\sqrt{5}a$, and its lattice structure is given in Fig. \ref{fig:lattice}.
In this lattice structure, we consider the tight-binding model with a nearest-neighbor in-plane hopping $t$ and interlayer hopping $t_z$ between two vertically aligned atoms labeled by $1_\pm$.
Then, the matrix elements of the Hamiltonian in the normal state $\hat{H}_n$ are as follows.
\begin{align}
&\langle i,j,1_\pm|H|i,j,2_\pm\rangle=
\langle i,j,2_\pm|H|i+1,j,3_\pm\rangle\nonumber\\
=&\langle i,j,3_\pm|H|i,j\pm1,4_\pm\rangle=
\langle i,j,4_\pm|H|i+1,j,5_\pm\rangle\nonumber\\
=&\langle i,j,5_\pm|H|i,j,1_\pm\rangle=-t,\\
&\langle i,j,1_\pm|H|i,j,3_\pm\rangle=
\langle i,j,2_\pm|H|1,j\pm1,4_\pm\rangle\nonumber\\
=&\langle i,j,3_\pm|H|i,j\pm1,5_\pm\rangle=
\langle i,j,4_\pm|H|i,j,1_\pm\rangle\nonumber\\
=&\langle i,j,5_\pm|H|i-1,j,2_\pm\rangle=-t,\\
&\langle i,j,1_+|H|i,j,1_-\rangle=-t_z,
\end{align}
and their Hermitian conjugates, where $|i,j,n_\pm\rangle$ is the basis ket of the $n$th
atom on the top ($n_+$) and the bottom ($n_-$) layer in the unit cell labeled by $i$ and $j$.
$i$ and $j$ are integers that specify the position of the unit cell along the $x$- and $y$-directions, respectively.
The Fourier-transformed form of the Hamiltonian
$\hat H_n({\bm k})$ is
\begin{align}
\hat H_n({\bm k})&=
\begin{pmatrix}
\hat H_{n}^+({\bm k})&\hat T_z({\bm k})\\
\hat T_z({\bm k})^\dagger&\hat H_{n}^-({\bm k})
\end{pmatrix},\label{eq:hn}\\
\hat H_{n}^\pm({\bm k})&\!=\!-t
{\renewcommand{\arraycolsep}{0pt}
\begin{pmatrix}
\mu/t &e^{ik_{1\pm} b}&e^{ik_{2\pm} b}& e^{-ik_{2\pm} b}& e^{-ik_{1\pm} b}\\
e^{-ik_{1\pm} b}&\mu/t&e^{ik_{1\pm} b}&e^{ik_{2\pm} b}& e^{-ik_{2\pm} b}\\
e^{-ik_{2\pm} b}&e^{-ik_{1\pm} b}&\mu/t&e^{ik_{1\pm} b}&e^{ik_{2\pm} b}\\
e^{ik_{2\pm} b}&e^{-ik_{2\pm} b}&e^{-ik_{1\pm} b}&\mu/t&e^{ik_{1\pm} b}\\
e^{ik_{1\pm} b}&e^{ik_{2\pm} b}&e^{-ik_{2\pm} b}&e^{-ik_{1\pm} b}&\mu/t\\
\end{pmatrix}},\label{eq:hnpm}\\
\hat T_{z}({\bm k})&=
{\renewcommand{\arraycolsep}{2pt}
\begin{pmatrix}
-t_z &0&0&0&0\\
0&0&0&0&0\\
0&0&0&0&0\\
0&0&0&0&0\\
0&0&0&0&0
\end{pmatrix}},\label{eq:tz}
\end{align}%
where $\mu$ is the chemical potential. $k_{1\pm}$ and $k_{2\pm}$ are given by $k_{1\pm}=(2k_x\pm k_y)/5$ and $k_{2\pm}=(-k_x\pm 2k_y)/5$.
In Eqs. (\ref{eq:hn}) to (\ref{eq:tz}), $\hat H_n^+({\bm k})$, $\hat H_n^{-}({\bm k})$, $\hat T_z({\bm k})$ represent the Hamiltonian for the top and the bottom layer, and the coupling between them, respectively.
It is useful to introduce the partially diagonalized basis which diagonalizes $\hat H_{n\pm}({\bm k})$,
\begin{align}
\hat U^\dagger\hat H^{\pm}_{n}({\bm k})\hat U&\nonumber\\
&\!\!\!\!\!\!\!\!\!\!\!\!\!\!\!\!\!\!\!\!\!\!\!=
\operatorname{diag}(
\varepsilon_{0}^\pm({\bm k}),
\varepsilon_{1}^\pm({\bm k}),
\varepsilon_{2}^\pm({\bm k}),
\varepsilon_{3}^\pm({\bm k}),
\varepsilon_{4}^\pm({\bm k})
),
\end{align}
\begin{align}
\varepsilon^{\pm}_{m}\!({\bm k})
\!=\!-2t\cos(k_{1\pm}b\!+\!2\pi m/5)\!-\!2t\cos(k_{2\pm}b\!+\!4\pi m/5)\!-\!\mu,
\end{align}
where the $(i,j)$-th element of $\hat U$ is given by $U_{ij}=\zeta_5^{(i-1)(j-1)}/\sqrt{5}$ with $\zeta_5=e^{2\pi\mathrm{i}/5}$ being the primitive 5th root of unity. In this partially diagonalized (PD) basis, all the elements of $\hat U^\dagger\hat T_{z}({\bm k})\hat U$ are equal to $-t_z/5\equiv -\tilde t_z$. Next, we consider the BdG Hamiltonian for superconducting states. The BdG Hamiltonian is represented by
\begin{align}
\hat H_s({\bm k})&=
\begin{pmatrix}
\hat H_{n}({\bm k})&\hat \Delta({\bm k})\\
\hat \Delta({\bm k})^\dagger&-\hat H_{n}({-\bm k})^t
\end{pmatrix},
\end{align}
where $\hat H_{n}({\bm k})$ is the normal state Hamiltonian given in Eq. (\ref{eq:hn}) and $\hat \Delta({\bm k})$ is the pair potential.
In this study, we consider an in-plane $d$-wave pairing. 
\begin{align}
\hat \Delta({\bm k})&=
\begin{pmatrix}
\hat \Delta_{+}({\bm k})&0\\
0&\hat \Delta_{-}({\bm k})
\end{pmatrix},\\
\hat U^\dagger\hat \Delta_{\pm}({\bm k})\hat U&=\mathrm{diag}\bigl(\Delta_{0}^\pm(\bm{k}), \allowbreak \Delta_{1}^\pm(\bm{k}), \allowbreak \dots, \allowbreak \Delta_{4}^\pm(\bm{k})\bigr),
\end{align}
\begin{align}
\Delta^{\pm}_{m}\!({\bm k})
\!&=\!\Delta_{\pm}\cos(k_{1\pm}b\!+\!2\pi m/5)\!-\!\Delta_\pm\cos(k_{2\pm}b\!+\!4\pi m/5)\!,
\end{align}
where $\Delta_{\pm}$ denotes the pair potential on the top and bottom layers.

\textit{Energy gap structure; nodal creation and flattening of BdG bands}.---By the diagonalization of the Hamiltonian in Eq. (9), we calculate the quasiparticle-excitation energy.
Here, we choose the parameters $\mu=-2.5 t$ and $\eta^\pm=10^{-3}t$.

Fig. 2 (a) shows the Fermi surfaces (FSs) in the case of $t_z=0.5t$. There are two FS, which are bonding and anti-bonding states of the bands for the top and the bottom layers. The obtained energy gap structure for $t_z=0, 0.02t$ and $0.5t$ are given in Fig. 2(b), (c) and (d), respectively, where $E_g$ is the lowest positive eigenvalue of energy, and  
$\varphi$ denotes the direction of the momentum as given in Fig. 2(a).
Without the interlayer hopping $t_z$, the energy gap structure shows a single-layer $d$-wave gap structure~\cite{SigristUeda,Kashiwaya_2000} and the nodal points appear at $\varphi=\pi/4\pm\theta/2$. In contrast, new nodal points appear at $\varphi=\pi/4$ when we include the interlayer hopping $t_z$ as shown in Fig 2(c) and (d). As shown in Fig. 3(a), the energy gap at $\varphi=\pi/4$ decreases with $t_z$ and the nodes are actually created when $|t_z|\gtrsim 0.6|\eta^\pm|$.
\begin{figure}[t!]
    \centering
    \includegraphics[width=8.6cm]{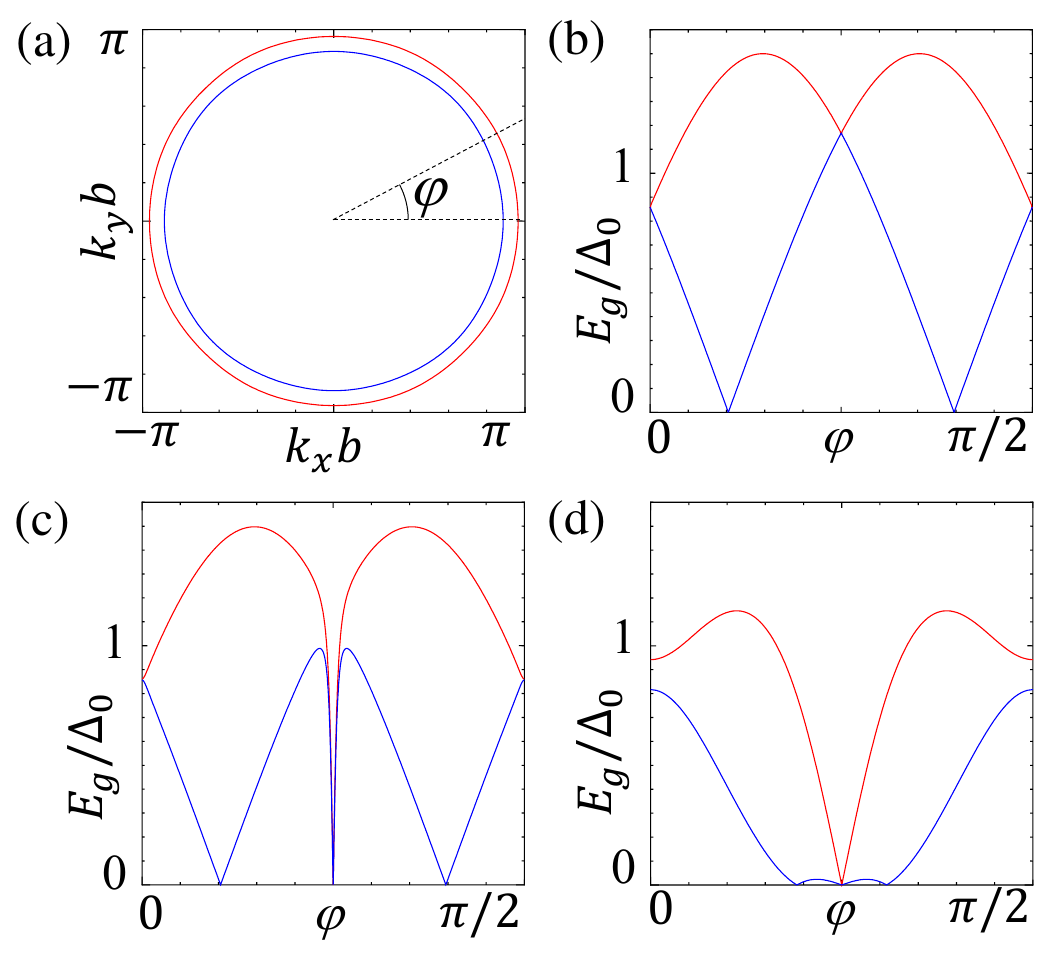}
    \caption{
    (a) FSs of the lowest (red) and the second-lowest bands (blue) at $\mu=-0.25t$ and $t_z=0.5t$. Angle dependence of the energy gaps $E_{g}$ of the lowest (red) and the second-lowest bands (blue) at (b) $t_z=0$, (c) $t_z=0.02t$, and (d) $t_z=0.5t$.}
    \label{fig:fseg}
\end{figure}%
\begin{figure}[t!]
    \centering
    \includegraphics[width=8.6cm]{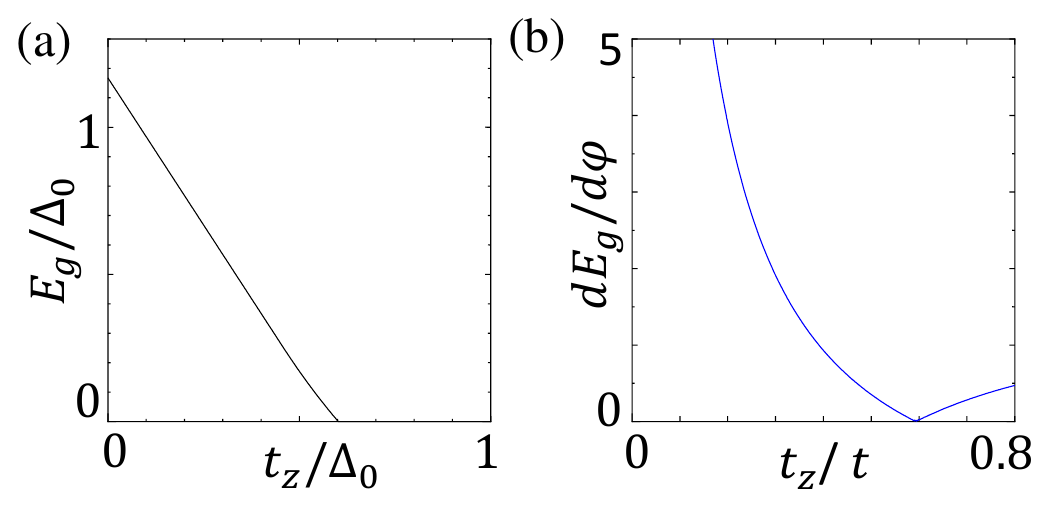}
    \caption{(a) $t_z$-dependence of the energy gap $E_g$ at $\varphi=\pi/4$. (b) The magnitude of the derivative of $E_g$ with respect to $\varphi$ at $\varphi=\pi/4$.  It is proportional to the group velocity along the FS.}
    \label{fig3}
\end{figure}%
\begin{figure}[t!]
    \centering
    \includegraphics[width=8.6cm]{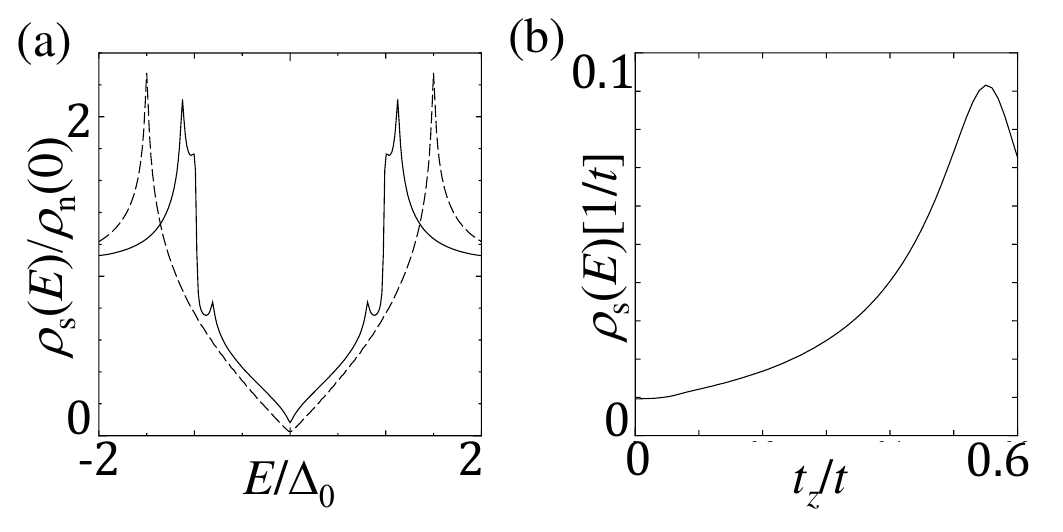}
    \caption{(a) The quasiparticle density of states $\rho_{S}(E)$ at $t_z=0$ (dashed line) and $t_z=0.6t$ (solid line). (b) $\rho_{S}(E)$ at $E=0$ is plotted as a function of $t_z$.}
    \label{fig4}
\end{figure}%

As we further increase interlayer hopping, we find that the position of original nodes on $\varphi=\pi/4\pm\theta/2$ gradually approaches $\varphi=\pi/4$. Simultaneously, the energy gap for the inner Fermi-surface band between the nodes is strongly suppressed. This suppression is characterized by the derivative of 
$E_g$ with respect to 
$\varphi$ shown in Fig 3(b).
Interestingly, the group velocity at $\varphi=\pi/4$ along the tangential direction to the FS, which is $\mathbf{n}=(1, -1)/\sqrt{2}$, becomes zero at $t_z\sim0.6t$.
This flattening of BdG quasiparticle bands reminds us of the flat bands in twisted bilayer graphene. Single-layer graphene and $d$-wave superconductor are similar in that they both exhibit band crossing points, which are protected by the winding number.
Other than this, we find that they both show the strong suppression of the group velocity in twisted bilayer systems. We refer to this flattened band as the Bogoliubov flat band in analogy with the flat band in twisted bilayer graphene.
In contrast to the flat band in twisted bilayer graphene, the flat band in twisted $d$-wave superconducting bilayer appears only along the direction perpendicular to the $\mathrm{C}_2$ rotation axis, which corresponds to the quasiparticle excitation by superconductivity.
To investigate the influence of the Bogoliubov flat band, we examine the density of states (DOS). As shown in Fig 4.(a), $\rho_s(E)$ has V-shaped structure at $E=0$ in both cases with and without $t_z$ because of the existence of nodes, where $\rho_s(E)$ is defined as
\begin{align}
\rho_s(E)=-\frac{1}{2\pi}\sum_{j}\int_\mathrm{BZ} \frac{d^2k}{(2\pi)^2}\mathrm{Im}\frac{1}{E-\xi_j({\bm k})+i\delta},
\end{align}
where $\xi_j({\bm k})$ is the $j$-th eigenvalue of the Hamiltonian matrix, and $\delta$ is infinitesimal number.
The V-shaped minimum is lifted by $t_z$ and $\rho_s(E=0)$ takes maximum at $t_z\sim0.55t$
This value of $t_z$ 
does not correspond to the value which minimizes the group velocity. This is because the strong suppression of the superconducting gap occurs between the nodes, not on the $\mathrm{C}_2$ rotation axis, and the position of these nodes 
gradually approaches the 
axis as $t_z$ increases.
Then, the range of $\varphi$ with an extremely small energy gap, which contributes to the zero energy DOS, becomes short.

\textit{Mechanism of Bogoliubov flat bands}.---We have shown that the interlayer hopping $t_z$ creates the nodes on the $\mathrm{C}_2$ rotation axes, and it also causes the strong suppression of the group velocity at the nodes.
To discuss the mechanism for the emergence of the Bogoliubov flat band, we introduce a $4\times 4$ effective Hamiltonian for simplicity in the following.
In the PD basis, $\hat H^{\pm}_{n}({\bm k})$ is diagonalized, and some of their components are away from the Fermi level. Thus, we consider the subspace, which is responsible for the low-energy quasiparticle excitations. In the present case with $\mu=-2.5t$, only $\varepsilon_0^\pm(\bm{k})$ is relevant to it.
Then, we consider the effective low-energy model,
\begin{align}
H_{\mathrm {eff}}({\bm k})=&
\begin{pmatrix}
    \varepsilon_0^+({\bm k}) & -\tilde t_z & \Delta_0^+({\bm k}) & 0\\
    -\tilde t_z & \varepsilon_0^-({\bm k}) & 0& \Delta_0^-({\bm k}) \\
    \Delta_0^+({\bm k}) & 0 & -\varepsilon_0^+({\bm k}) & \tilde t_z\\
    0& \Delta_0^-({\bm k}) & \tilde t_z & -\varepsilon_0^-({\bm k})
\end{pmatrix}\nonumber\\
=&\varepsilon^s({\bm k})\sigma_0\tau_3+\varepsilon^a({\bm k})\sigma_3\tau_3+\Delta^s({\bm k})\sigma_0\tau_1\nonumber\\&
+\Delta^a({\bm k})\sigma_3\tau_1-\tilde t_z\sigma_1\tau_3, \label{eq:Heff}
\end{align}
where $2\varepsilon^{s,a}({\bm k})=\varepsilon_0^+({\bm k})\pm\varepsilon_0^-({\bm k})$ and 
$2\Delta^{s,a}({\bm k})=\Delta_0^+({\bm k})\pm\Delta_0^-({\bm k})$.
Along the diagonal direction $k_x=\pm k_y$, $\varepsilon_0^+({\bm k})=\varepsilon_0^-({\bm k})$ because of the $\mathrm{C}_2$ rotational symmetry and $\varepsilon^a({\bm k})=0$. Likely, in the present case with $\eta^+=\eta^-$, $\Delta^s({\bm k})=0$ at $k_x=\pm k_y$.
Since the second and the third terms are absent along the diagonal direction, we can easily diagonalize the effective Hamiltonian by the unitary transformation $U=\exp(-i\frac{\pi}{4}\sigma_2\tau_0)\exp(i\frac{\theta}{2}\sigma_1\tau_2)$ with $\cos\theta=\varepsilon^s({\bm k})/\sqrt{\varepsilon^s({\bm k})^2+\Delta^a({\bm k})^2}$ and $\sin\theta=\Delta^a({\bm k})/\sqrt{\varepsilon^s({\bm k})^2+\Delta^a({\bm k})^2}$,
\begin{align}
U^\dagger H_{\mathrm {eff}}({\bm k})U=&
\sqrt{\varepsilon^s({\bm k})^2+\Delta^a({\bm k})^2}\sigma_0\tau_3-\tilde t_z\sigma_3\tau_3, \label{eq:uHeff}
\end{align}
for $k_x=k_y$,
and we obtain the eigenvalues,
\begin{align}
E=\pm\left( \sqrt{\varepsilon^s({\bm k})^2+\Delta^a({\bm k})^2}\pm \tilde t_z\right).
\end{align}
It follows from this equation that there exist zero-energy eigenvalues when $\tilde t_z^2=\varepsilon^s({\bm k})^2+\Delta^a({\bm k})^2$ is satisfied. Actually, the nodal points are formed when $\tilde t_z\ge |\Delta^a({\bm k_\mathrm {F}})|$,
where $\bm k_\mathrm {F}$ is the Fermi wave vector along the diagonal direction. 
Next, we discuss the reduction of the group velocity.
For that purpose, we consider 
the first-order change in the energy eigenvalue with respect to a shift in the wave vector from the node,
\begin{align}
H_{\mathrm {eff}}({\bm k}_n+\delta{\bm k})=&H_{\mathrm {eff}}({\bm k}_n)+\delta H_{\mathrm {eff}}({\bm k}_n),
\end{align}
where we take 
the shift of the wave vector $\delta{\bm k}$ to be perpendicular to $\mathrm{C}_2$ rotation axis, $\delta{\bm k}=\delta k\mathbf{n}$ with directional unit vector $\mathbf{n}=(-1, 1)/\sqrt{2}$. Based on the first-order perturbation for the degenerate zero-energy eigenvalues, we obtain the group velocity in the direction of $\mathbf{n}$ as 
\begin{align}
&E_\pm({\bm k}_n+\delta k\mathbf{n})\nonumber\\&=\pm \left|
\frac{
\Delta^a_0({\bm k_n})\mathbf{n}\cdot \nabla \varepsilon^a_{mm'}({\bm k}_n)
-\varepsilon^s_0({\bm k}_n)\mathbf{n}\cdot\nabla \Delta^s_{mm'}({\bm k}_n)}{\sqrt{\varepsilon^s_0({\bm k}_n)^2+\Delta^a_0({\bm k_n})^2}}
\right|\delta k\nonumber\\
&=\pm\sqrt{\varepsilon^+_0({\bm k}_n)^2+\Delta^+_0({\bm k_n})^2}\left|\mathbf{n}\cdot\nabla\phi({\bm k}_n)
\right|\delta k
\nonumber\\
&\equiv \pm v_{{\bm k}_n}\delta k,\end{align}
where $\phi({\bm k}_n)$ is the phase of complex number $\varepsilon^+_0({\bm k}_n)+i\Delta^+_0({\bm k_n})=\sqrt{\varepsilon^+_0({\bm k}_n)^2+\Delta^+_0({\bm k_n})^2}\exp(i\phi({\bm k}_n))$. The details of the derivation of Eq. (18) are given in the supplementary material. It follows from Eq. (18) that the group velocity becomes zero when the gradient of the phase at the nodal point $\phi({\bm k}_n)$ is parallel to the $\mathrm{C}_2$ rotation axis. Interestingly, half of $\nabla\phi({\bm k}_n)$ is the Berry connection $\mathbf{A}({\bm k})$ for single layer system of top layer, and it relates to the winding number $W$ by $W=\oint_C \mathbf{A}({\bm k})\cdot d{\bm k}/\pi$\cite{TAndo1998,Mikitik1999,Ryu_2010}. When we choose a closed curve $C$ enclosing the nodal point without $t_z$, which is the phase singularity, $|W|$ becomes unity. Thus, the the direction of $\mathbf{A}({\bm k})$ rotate $-2\pi$ around the nodes of the top layer as shown by the arrows in Fig. 5. From this nodal point, we can write the line where the direction of $\mathbf{A}({\bm k})$ is $-3/4\pi$ by red line. When a nodal point appears at the crossing point of this red line and the $\mathrm{C}_2$ rotation axis $k_x=k_y$ the group velocity becomes zero because $\mathbf{A}({\bm k})$ is parallel to the rotation axis and $\mathbf{n}\cdot \mathbf{A}({\bm k})$ is zero. This can be realized by the changing Fermi surface by introducing $t_z$ as shown in blue solid lined in Fig. 5.
In general, the closer the phase singularity point is to the 
$\mathrm{C}_2$ rotation axis, the smaller the value of 
$t_z$ required to realize the Bogoliubov flat band.
Even if the the condition described above is not satisfied for given single layer Hamiltonian and corresponding Berry connection $\mathbf{A}({\bm k})$ and $t_z$, Bogoliubov flat band can be realized by changing the twisting angle.
\begin{figure}[t!]
    \centering
    \includegraphics[width=\linewidth]{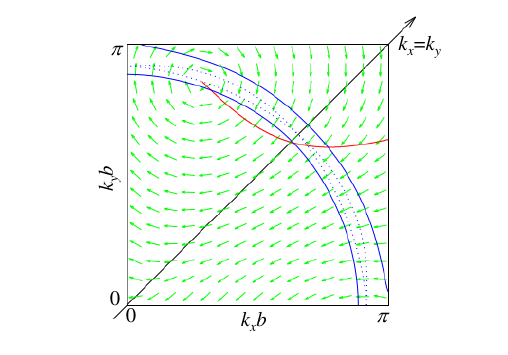}
    \caption{
    Illustration of a nodal creation with zero group velocity in ${\bm k}$-space. The blue solid and dotted lines denote the FSs with and without interlayer hopping, respectively. The arrows show the direction of the Berry connection of the top layer, $\mathbf{A}({\bm k})/|\mathbf{A}({\bm k})|$. At the momentum on the Red line, the direction of the Berry connection is $-3/4\pi$. 
    }
    \label{fig5}
\end{figure}

We have studied twisted bilayer systems composed of nodal $d$-wave superconductors. By solving the Bogoliubov de Gennes equation, we find the emergence of the nontrivial nodes on the $\mathrm{C}_2$ rotation axis and the Bogoliubov flat bands near the nodes. 
By the analysis of $4 \time 4$  effective Hamiltonian, we find the condition to realize Bogoliubov bands; the Berry connection of the single layer system is parallel to the $\mathrm{C}_2$ rotation axis at the nontrivial node. Our findings establish a new paradigm: superconducting twistronics, where the twist angle serves as a tuning knob to design gapless flat band superconductors and opens a pathway toward engineering novel quantum phases in twisted superconducting heterostructures. 

In this work, we focus on the energy spectrum of a twisting bilayer superconductor with the pairing, which is odd under the in-plane $\mathrm{C}_2$ rotation. It should be examined
whether this pairing is a stable solution of the gap equation.
We outline the following three directions for future research:
The first one is to elucidate the relation between Bogoliubov flat band and quantum metric because quantum geometric part of superfluid weight becomes dominant in flat band system\cite{Peotta2015}.
Second, it would be important to study the resulting odd-frequency pairing since it is known that gapless superconducting states inevitably accompany odd-frequency pairing \cite{Berezinskii,tanaka12,Tanaka2007PRL,tanaka12,tanaka2024theory, LinderBalatsky,Cayao2020}. Since the generation of odd-frequency pairing has been discussed in multi-orbital system including flat band systems  \cite{BlackShaffer,Ankita}, the relevance of the odd-frequency pairing and Bogoliubov flat band should be clarified. 
Finally, it is not yet understood the effect of the Bogoliubov flat band on the zero energy edge states in $d$-wave superconductors\cite{HuPRL1994,Tanaka95,Kashiwaya_2000,
Lofwander,SatoPRB2011}.
\begin{acknowledgements}
K.\ Y.\ and Y.\ F.\ acknowledge financial support from the Sumitomo Foundation.
K.\ Y., Y.\ F., and Y.\ T.\ acknowledge financial support from JSPS with Grants-in-Aid for Scientific Research (KAKENHI Grants No. 25K07203).  
Y. T. acknowledges financial support from JSPS with Grants-in-Aid for Scientific Research (KAKENHI Grants Nos.\ 23K17668, 24K00583, 24K00556, 24K00578, 25H00609, and 25H00613). 
\end{acknowledgements}

\bibliography{biblio}
\clearpage
\onecolumngrid

\appendix
\section*{Supplemental Material}

\setcounter{equation}{0}
\renewcommand{\theequation}{S\arabic{equation}}
\setcounter{figure}{0}
\renewcommand{\thefigure}{S\arabic{figure}}

\section{S1. Analysys of 4$\times$4 effective Hamiltonian}

In the main text, we have introduced the 4$\times4$  effective Hamiltonian for twisted layered superconductors.
In this section, we show the details of its derivation, creation of nodes, and the quasiparticle excitation around the nodes.

\subsection{S1.1: Derivation of the effective Hamiltonian}
\begin{figure}[t!]
    \centering
    \includegraphics[width=\linewidth]{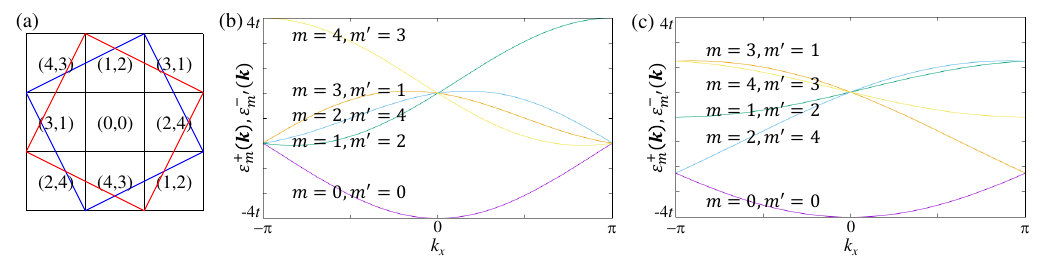}
    \caption{
    (a) The mapping between the original Brillouin zone and the folded Brillouin zone. The blue and red lines denote the original Brillouin zone for the top and bottom layer, respectively, and the black lines denote the extended Brillouin zone for the twisted bilayer system.  The numbers in parentheses denote the corresponding values of $m$ and $m'$ in each extended Brillouin zone. Diagonal components of the Hamiltonian $\varepsilon_m^\pm({\bm k})$ along (b) diagonal axis $k_x=k_y$ and (c) the $k_x$-axis.}
    \label{fig:smfig1}
\end{figure}

As given in the main text, the intralayer part of the normal-state Hamiltonian is diagonalized in the partially-diagonalized (PD) basis, and their diagonal components are given by
\begin{align}
\hat U^\dagger\hat H^{\pm}_{n}({\bm k})\hat U&=
\operatorname{diag}(
\varepsilon_{0}^\pm({\bm k}),
\varepsilon_{1}^\pm({\bm k}),
\varepsilon_{2}^\pm({\bm k}),
\varepsilon_{3}^\pm({\bm k}),
\varepsilon_{4}^\pm({\bm k})
),\\
\varepsilon^{\pm}_{m}({\bm k})
&=-2t\cos(k_{1}^{\pm}b+2\pi m/5)-2t\cos(k_{2}^{\pm}b+4\pi m/5)-\mu,\label{eq:diag}\\
\hat U^\dagger\hat \Delta^{\pm}({\bm k})\hat U&=
\operatorname{diag}(
\Delta_{0}^\pm({\bm k}),
\Delta_{1}^\pm({\bm k}),
\Delta_{2}^\pm({\bm k}),
\Delta_{3}^\pm({\bm k}),
\Delta_{4}^\pm({\bm k})
),\\
\Delta^{\pm}_{m}({\bm k})
&=2\eta^\pm\cos(k_{1}^{\pm}b+2\pi m/5)-2\eta^\pm\cos(k_{2}^{\pm}b+4\pi m/5),\label{eq:diag2}
\end{align}
where $U$, $\hat H^{\pm}_{n}({\bm k})$ and $\hat \Delta^{\pm}_{n}({\bm k})$ is given in the main text, $k_{1\pm}=(2k_x\pm k_y)/5$ and $k_{2\pm}=(-k_x\pm 2k_y)/5$. Since the nontrivial nodes appear on the $k_x$- and $k_y$-axes or the diagonal axes $k_x=\pm k_y$, we focus on the degeneracy of the diagonal component given in eq. (\ref{eq:diag}) and (\ref{eq:diag2}).
Along the diagonal axis $k_x=k_y$, two relations $k_{1}^+=-k_{2}^-$ and $k_{1}^-=k_{2}^+$ are satisfied, and therefore $\varepsilon_{m}^+({\bm k})=\varepsilon_{m'}^-({\bm k})$ where $m'\equiv 2m$ (mod 5).
Similarly, along the $k_x$-axis, $\varepsilon_{m}^+({\bm k})=\varepsilon_{m'}^-({\bm k})$ for $m=m'$.
As for the pair potential $\Delta_m^+({\bm k})/\eta^+=\Delta_m^-({\bm k})/\eta^-$ along the $k_x$-axis and $\Delta_m^+({\bm k})/\eta^+=-\Delta_{m'}^-({\bm k})/\eta^-$ along the diagonal axis $k_x=k_y$. As noted in the following, the gap nodes can be formed in the case with $\varepsilon_m^+({\bm k})=\varepsilon_{m'}^-({\bm k})$ and $\Delta_m^+({\bm k})=-\Delta_{m'}^-({\bm k})$. Thus, gap nodes appear along the diagonal axis $k_x=\pm k_y$ in the case of $\eta^+=\eta^-$ or on the $k_x$- and $k_y$-axis in the case of $\eta^+=\eta^-$.
As shown in the Fig. \ref{fig:smfig1},
$\varepsilon_m^\pm({\bm k})$ degenerate at the zone center or the zone boundary for different values of $m$. However, away from these points, $\varepsilon_m^\pm({\bm k})$ can be identified. 
On the other hand,
$\tilde t_z$ makes the hybridization between $\varepsilon_m^+({\bm k})$ and 
$\varepsilon_{m'}^-({\bm k})$ for any $m$ and $m'$ values. However, the effect of this hybridization becomes small in the case where the level splitting $|\varepsilon_m^+({\bm k})-\varepsilon_{m'}^-({\bm k})|$ is much smaller than $|\tilde t_z|$. Thus, we consider the $4\times 4$ effective Hamiltonian which is the subspace including degenerate pairs of $\varepsilon_{m}^+({\bm k})$ and $\varepsilon_{m'}^-({\bm k})$.  
\begin{align}
H^{\mathrm {eff}}_{mm'}({\bm k})=&
\begin{pmatrix}
    \varepsilon_m^+({\bm k}) & -\tilde t_z & \Delta_m^+({\bm k}) & 0\\
    -\tilde t_z & \varepsilon_{m'}^-({\bm k}) & 0& \Delta_{m'}^-({\bm k}) \\
    \Delta_m^+({\bm k}) & 0 & -\varepsilon_{m}^+({\bm k}) & \tilde t_z\\
    0& \Delta_{m'}^-({\bm k}) & \tilde t_z & -\varepsilon_{m'}^-({\bm k})
\end{pmatrix}\nonumber\\
=&\varepsilon^s_{mm'}({\bm k})\sigma_0\tau_3+\varepsilon^a_{mm'}({\bm k})\sigma_3\tau_3+\Delta^\mathrm{s}_{mm'}({\bm k})\sigma_0\tau_1+\Delta^\mathrm{a}_{mm'}({\bm k})\sigma_3\tau_1-\tilde t_z\sigma_1\tau_3, \label{eq:Heff}
\end{align}
where $\varepsilon^{s,a}_{mm'}({\bm k})=(\varepsilon^+_{m}({\bm k})\pm \varepsilon^-_{m'}({\bm k}))/2$, and 
$\Delta^{\mathrm{s,a}}_{mm'}({\bm k})=(\Delta^+_{m}({\bm k})\pm\Delta^-_{m'}({\bm k}))/2$.

\subsection{S1.2: Creation of the nodes}
The effective Hamiltonian introduced in the last section can be easily diagonalized in the case where the second and the third terms in Eq. (\ref{eq:Heff}) are absent. This happens on the diagonal axes $k_x=\pm k_y$ for $\eta^+=\eta^-$ or on the $k_x$- and $k_y$-axes for $\eta^+=-\eta^-$ because of $\mathrm{C}_2$ rotational symmetry on these axces.
In this case, the eigenvalues are given by $\pm\sqrt{\varepsilon^s_{mm'}({\bm k})^2+\Delta^a_{mm'}({\bm k})^2}\pm \tilde t_z$.
Thus, the nodal points appear at ${\bm k}$
which satisfies $\varepsilon^s_{mm'}({\bm k})^2+\Delta^a_{mm'}({\bm k})^2= \tilde t_z^2$. This wave vector almost corresponds to the Fermi wave vector, which satisfies $\varepsilon^s_{mm'}({\bm k})^2= \tilde t_z^2$ because the pair potential $|\eta^\pm|$ is much smaller than the hopping integral $|t|$ in most cases.

\subsection{S1.3: Quasiparticle excitations around the nodes}
In this section, we derive the group velocity at the nodes and discuss the flattening of the Bogoliubov bands.
For that purpose, we examine how the Hamiltonian changes when the wave vector is slightly varied from the nodal point.
Suppose the wave vector on the nodal point is given by ${\bm k}_n$, we set the shift of the wave vector $\delta {\bm k}=\delta k\mathbf{n}$ where $\delta k$ and $\mathbf{n}$ describe the amount of the shift and the unit vector specifying the direction of the shift, respectively.
Since we want to discuss the group velocity along the Fermi surface, we set $\mathbf{n}=(-1, 1)\sqrt{2}$ for the nodes on the diagonal axes $k_x=k_y$, and 
$\mathbf{n}=(0, 1)$.
Since the system has $\mathrm{C}_4$ rotational symmetry along the $z$-axis, the group velocity at the nodes on the $k_x=-k_y$ and $k_y$-axis is the same as in the case mentioned above.
As discussed in the previous section, $\varepsilon^a_{mm'}({\bm k})$ and $\Delta^s_{mm'}({\bm k})$ are absent at the nodal point ${\bm k}={\bm k}_n$,
the Hamiltonian on the node is given by
\begin{align}
H^{\mathrm {eff}}_{mm'}({\bm k}_n)=\varepsilon^s_{mm'}({\bm k}_n)\sigma_0\tau_3
+\Delta^\mathrm{a}_{mm'}({\bm k}_n)\sigma_3\tau_1-\tilde t_z\sigma_1\tau_3.
\end{align}
By the small shift of the wave vector,
the Hamiltonian is given by
\begin{align}
H^{\mathrm {eff}}_{mm'}({\bm k}_n+\delta{\bm k})=H^{\mathrm {eff}}_{mm'}({\bm k}_n)+\delta H^{\mathrm {eff}}_{mm'}({\bm k}_n).
\end{align}
where $\delta H^{\mathrm {eff}}_{mm'}({\bm k}_n)$ is considered up to first order in $\delta k$ since the higher order terms do not affect the group velocity,
\begin{align}
\delta H^{\mathrm {eff}}_{mm'}({\bm k}_n)=\mathbf{n}\cdot \left.\nabla H^{\mathrm {eff}}_{mm'}({\bm k})\right|_{{\bm k}={\bm k}_n} \delta k.
\end{align}
It is found that only $\varepsilon^a_{mm'}({\bm k})$ and $\Delta^\mathrm{s}_{mm'}({\bm k})$ are relevant to this first-order change of the Hamiltonian.
More precisely, $\varepsilon^a_{mm'}(k_x, k_y)=-\varepsilon^a_{mm'}(k_y, k_x)=-\varepsilon^a_{mm'}(k_x, -k_y)$ for degenerate pairs of $m$ and $m'$. As for the pair potential, $\Delta^s_{mm'}(k_x, k_y)=-\Delta^s_{mm'}(k_y, k_x)=\Delta^s_{mm'}(k_x, -k_y)$ for $\eta^+=\eta^-$, and $\Delta^s_{mm'}(k_x, k_y)=\Delta^s_{mm'}(k_y, k_x)=-\Delta^s_{mm'}(k_x, -k_y)$ for $\eta^+=-\eta^-$. Their relations are owing to the $\mathrm{C}_2$ rotational symmetry at $k_x=0$ and $k_x=k_y$.
Thus, when we consider the change of the Hamiltonian by the shift of the wave vector along $\mathbf{n}$, $\varepsilon^a_{mm'}({\bm k})$ and $\Delta^\mathrm{s}_{mm'}({\bm k})$ are antisymmetric function and the other terms are symmetric. Then,
\begin{align}
\delta H^{\mathrm {eff}}_{mm'}({\bm k}_n)=
\mathbf{n}\cdot \left.\nabla \varepsilon^a_{mm'}({\bm k})\right|_{{\bm k}={\bm k}_n} \delta k\sigma_3\tau_3
+\mathbf{n}\cdot \left.\nabla \Delta^\mathrm{s}_{mm'}({\bm k})\right|_{{\bm k}={\bm k}_n} \delta k\sigma_0\tau_1.
\end{align}
Next, we examine the change of the eigenvalues by $\delta H^{\mathrm {eff}}_{mm'}({\bm k}_n)$. For that purpose, we change the basis which diagonalizes $ H^{\mathrm {eff}}_{mm'}({\bm k}_n)$. $ H^{\mathrm {eff}}_{mm'}({\bm k}_n)$ is diagonalized by the unitary transformation
$U_{{\bm k}_n}=\exp(-i\frac{\pi}{4}\sigma_2\tau_0)\exp(i\frac{\theta_{{\bm k}_n}}{2}\sigma_1\tau_2)$ with $\cos\theta_{{\bm k}_n}=\varepsilon^s({{\bm k}_n})/\sqrt{\varepsilon^s({{\bm k}_n})^2+\Delta^a({{\bm k}_n})^2}$ and $\sin\theta_{\bm k}=\Delta^a({{\bm k}_n})/\sqrt{\varepsilon^s({{\bm k}_n})^2+\Delta^a({{\bm k}_n})^2}$.
Then, the effective Hamiltonian around the nodes is transformed as
\begin{align}
U_{{\bm k}_n}^\dagger H^{\mathrm {eff}}_{mm'}({\bm k}_n)U_{{\bm k}_n}
&=
\sqrt{\varepsilon^s_{mm'}({{\bm k}_n})^2+\Delta^a_{mm'}({\bm k}_n)^2}\sigma_0\tau_3-\tilde t_z\sigma_3\tau_3\label{eq:s10}
\\
U_{{\bm k}_n}^\dagger\delta H^{\mathrm {eff}}_{mm'}({\bm k}_n)U_{{\bm k}_n}
&=
-\delta k\left(\sin\theta_{{\bm k}_n}\mathbf{n}\cdot \left.\nabla \varepsilon^a_{mm'}({\bm k})\right|_{{\bm k}={\bm k}_n}
-\cos\theta_{{\bm k}_n}\mathbf{n}\cdot \left.\nabla \Delta^s_{mm'}({\bm k})\right|_{{\bm k}={\bm k}_n}
\right)\sigma_0\tau_1
\nonumber\\
&\quad -\delta k\left(\cos\theta_{{\bm k}_n}\mathbf{n}\cdot \left.\nabla \varepsilon^a_{mm'}({\bm k})\right|_{{\bm k}={\bm k}_n}
+\sin\theta_{{\bm k}_n}\mathbf{n}\cdot \left.\nabla \Delta^s_{mm'}({\bm k})\right|_{{\bm k}={\bm k}_n}
\right)\sigma_1\tau_3.\label{eq:s11}
\end{align}
While Eq. (\ref{eq:s11}) does not have any diagonal components, it may give the first-order change in the energy eigenvalue with respect to $\delta k$ when degeneracy occurs in the energy eigenvalues in Eq. (\ref{eq:s10}), i.e., zero-energy states of the nodes.
Here, we consider the case with $\tilde t_z>0$. In this case, nodes in Eq. (\ref{eq:s10}) appear in the subspace of $\sigma_3=+1$. The effective Hamiltonian in this subspace takes the form, 
\begin{align}
\langle \sigma_3=+1|U_{{\bm k}_n}^\dagger H^{\mathrm {eff}}_{mm'}({\bm k}_n)U_{{\bm k}_n}|\sigma_3=+1\rangle
&=
\left(\sqrt{\varepsilon^s_{mm'}({{\bm k}_n})^2+\Delta^a_{mm'}({\bm k}_n)^2}-\tilde t_z\right)\tau_3\label{eq:s12}
\\
\langle \sigma_3=+1|U_{{\bm k}_n}^\dagger\delta H^{\mathrm {eff}}_{mm'}({\bm k}_n)U_{{\bm k}_n}|\sigma_3=+1\rangle
&=
-\delta k\left(\sin\theta_{{\bm k}_n}\mathbf{n}\cdot \left.\nabla \varepsilon^a_{mm'}({\bm k})\right|_{{\bm k}={\bm k}_n}
-\cos\theta_{{\bm k}_n}\mathbf{n}\cdot \left.\nabla \Delta^s_{mm'}({\bm k})\right|_{{\bm k}={\bm k}_n}
\right)\tau_1.\label{eq:s13}
\end{align}
Since we choose ${\bm k}_n$ such that $\sqrt{\varepsilon^s_{mm'}({{\bm k}_n})^2+\Delta^a_{mm'}({\bm k}_n)^2}-\tilde t_z$, Eq. (\ref{eq:s12}) is absent. Then, the energy eigenvalues at ${\bm k}={\bm k}_n+\delta{\bm k}$ is given by
\begin{align}
E_\pm({\bm k}_n+\delta{\bm k})&=\pm \delta k\left|\sin\theta_{{\bm k}_n}\mathbf{n}\cdot \left.\nabla \varepsilon^a_{mm'}({\bm k})\right|_{{\bm k}={\bm k}_n}
-\cos\theta_{{\bm k}_n}\mathbf{n}\cdot \left.\nabla \Delta^s_{mm'}({\bm k})\right|_{{\bm k}={\bm k}_n}
\right|\nonumber\\
&\equiv \pm v_{{\bm k}_n}\delta k,\label{eq:s14}
\end{align}
where $v_{{\bm k}_n}$ is the group velocity at the nodes given by
\begin{align}
v_{{\bm k}_n}=\lim_{\delta k \to 0} \frac{E_+({\bm k}_n+\delta{\bm k})-E_+({\bm k}_n)}{\delta k}.
\end{align}
Note that the effective Hamiltonian and resultant group velocity in the subspace of $\sigma_3=-1$, which is available in the case $\tilde t_z<0$, are the same because Eq. (\ref{eq:s13}) has the same form in this subspace. Next, we examine the explicit form of $\pm v_{{\bm k}_n}$ in the case $m=m'=0$, $\eta^+=\eta^-$ and $\mathbf{n}=(-1, 1)/\sqrt{2}$, which were considered in the main text.
From Eq. (\ref{eq:s14}), we obtain
\begin{align}
v_{{\bm k}_n}=\frac{1}{\sqrt{\varepsilon^s_{00}({{\bm k}_n})^2+\Delta^a_{00}({\bm k}_n)^2}}
\left|
\varepsilon^s_{00}({\bm k}_n)\left(
\frac{d\Delta^s_{00}({{\bm k}_n})}{dk_x}
-\frac{d\Delta^s_{00}({{\bm k}_n})}{dk_y}
\right)
-\Delta^a_{00}({\bm k}_n)\left(
\frac{d\varepsilon^a_{00}({{\bm k}_n})}{dk_x}
-\frac{d\varepsilon^a_{00}({{\bm k}_n})}{dk_y}
\right)
\right|.\label{eq:s16}
\end{align}
Since the nodal point is on the
$\mathrm{C}_2$ rotation axis $k_x=k_y$,
we have the relations $\varepsilon_0^+(k_x, k_y)=\varepsilon_0^-(k_y, k_x)$ and
$\Delta_0^+(k_x, k_y)=-\Delta_0^-(k_y, k_x)$. Then Eq. (\ref{eq:s16}) becomes
\begin{align}
v_{{\bm k}_n}&=\frac{1}{2\sqrt{\varepsilon^+_{0}({{\bm k}_n})^2+\Delta^+_{0}({\bm k}_n)^2}}
\left|
\varepsilon^+_{0}({\bm k}_n)\left(
\frac{d\Delta^+_{0}({{\bm k}_n})}{dk_x}
-\frac{d\Delta^+_{0}({{\bm k}_n})}{dk_y}
\right)
-\Delta^+_{0}({\bm k}_n)\left(
\frac{d\varepsilon^+_{0}({{\bm k}_n})}{dk_x}
-\frac{d\varepsilon^+_{0}({{\bm k}_n})}{dk_y}
\right)
\right|\nonumber\\
&=\frac{1}{\sqrt{2}}\left|\frac{d\phi({\bm k}_n)}{dk_x}-\frac{d\phi({\bm k}_n)}{dk_y}\right|,\label{eq:s17}
\end{align}
where $\phi({\bm k})$ is the phase of $\varepsilon_0^+({\bm k})+i\Delta_0^+({\bm k})\equiv
\sqrt{\varepsilon_0^+({\bm k})^2+\Delta_0^+({\bm k})^2}\exp(i\phi({\bm k}))$. This result shows that the group velocity at the nodes along the direction perpendicular to ${\bm k}_n$ corresponds to the gradient of $\phi({\bm k})$ to that direction. As seen from Eq. (\ref{eq:s17}), the group velocity has no explicit dependence on interlayer hopping integral $\tilde t_z$. However, ${\bm k}_n$ changes with $\tilde t_z$, because ${\bm k}_n$ is determined to satisfy $\varepsilon_0^+({\bm k})^2+\Delta_0^+({\bm k})^2={\tilde t_z}^2$. This is the reason why the group velocity actually depends on $\tilde t_z$.

\end{document}